\documentclass[aps,english,showpacs,10pt]{revtex4}
\usepackage[latin1]{inputenc}
\usepackage{epsfig}
\usepackage{float}
\usepackage{latexsym}
\usepackage{graphicx}
\usepackage{amssymb}
\usepackage{amsmath}
\usepackage{amsfonts}
\usepackage{hyperref}
\usepackage{color}
\allowdisplaybreaks[1]
\newcommand{\be}{\begin{equation}}
\newcommand{\ee}{\end{equation}}
\newcommand{\ben}{\begin{eqnarray}}
\newcommand{\een}{\end{eqnarray}}
\newcommand{\bes}{\begin{subequations}}
\newcommand{\ees}{\end{subequations}}
\newcommand{\bb}{\bibitem}

\newcommand{\nn}{\nonumber\\}
\newcommand{\re}{{\rm e}}
\newcommand{\Frac}[2]{\frac{{\displaystyle #1}}{{\displaystyle #2}}}
\begin{document}
\title{Exact cosmological solutions of models with an interacting dark
  sector }

\author{A. B. Pavan $^a$}\email{alan@unifei.edu.br} \author{Elisa
  G. M. Ferreira$^b$}\email{elisa@fma.if.usp.br} \author{Sandro
  M. R. Micheletti$^c$}\email{smrm@fma.if.usp.br} \author{J. C. C. de
  Souza$^d$}\email{jose.souza@ufabc.edu.br}
\author{E. Abdalla$^b$}\email{eabdalla@fma.if.usp.br}

\affiliation{$^a$Instituto de Ci\^{e}ncias Exatas, Universidade
  Federal de Itajub\'{a}, Avenida BPS 1303 Pinheirinho, 37500-903,
  Itajub\'{a}, MG, Brazil}

\affiliation{$^b$Instituto de F\'isica, Universidade de S\~ao Paulo,
  C.P. 66318, 05315-970, S\~ao Paulo, SP, Brazil}

\affiliation{$^c$Universidade Federal do Rio de Janeiro, Campus Macaé,
  Avenida Aluizio da Silva Gomes, 50, Granja dos Cavaleiros,
  27930-560, Macaé, Rio de Janeiro, Brazil}

\affiliation{$^d$Centro de Ciências Naturais e Humanas, Universidade
  Federal do ABC, Rua Santa Adélia 166, 09210-170, Santo André, São
  Paulo, Brazil}

\date{\today}
\begin{abstract}
{In this work we extend the First Order Formalism for cosmological
  models that present an interaction between a fermionic and a scalar
  field. Cosmological exact solutions describing universes filled with
  interacting dark energy and dark matter have been obtained. Viable
  cosmological solutions with an early period of decelerated expansion
  followed by late acceleration have been found, notably one which
  presents a dark matter component dominating in the past and a dark
  energy component dominating in the future. In another one, the dark
  energy alone is the responsible for both periods, similar to a
  Chaplygin gas case. Exclusively accelerating solutions have also
  been obtained.}
\end{abstract}


\maketitle

\section{Introduction}

The recent improvements in the observational techniques available for
the measurement of cosmological parameters show, with increasing
accuracy, that the universe is composed mainly of two mysterious
entities, the so-called dark energy and dark matter. The first one is
believed to be the cause of the observed accelerated expansion of the
universe \cite{a1,a2,a3}, and corresponds to approximatelly 70\% of
its total energy density. The latter corresponds to almost 25\% of the
energy density of the universe and plays an important role in large
structure formation. The reader is referred to \cite{pr}-\cite{rf} for
a review on the theoretical developments that followed these
observations.

An interaction between these two dark sectors is rather plausible, and
can be even considered to be a necessary feature in cosmological
models based on Quantum Field Theory. An extensive literature in the
last years has shown both theoretical and phenomenological aspects of
the coupling between dark matter and dark energy (see, e.g.,
\cite{wetterich}-\cite{clusters}). The present work is an attempt to
solve the equations of motion for a theoretical model based on an
interaction Lagrangian, in which dark energy is represented by a
canonical scalar field, and dark matter takes the form of a fermionic
field.

Solving these equations, even for the case in which no interaction is
taken into account, is very often an extremely arduous task, if not
totally impossible, without the use of certain approximations. Many
papers rely on intensive computer simulations to gather some insight
about cosmological parameters, while others tackle the difficulties by
analysing the phase space dynamics represented by the generic
equations of motion for the cosmological model in question.

In all cases, the viable cosmological solution including dark energy
must obey that the energy density of scalar field must remain
subdominant during radiation and matter dominant eras to allow for
structure formation, becoming dominant in late times accounting for
the current acceleration of the universe. This can be obtained by
\textit{scaling solutions} where the energy density of dark energy
mimics the background fluid (radiation or matter)
\cite{dynamic_DE}. The field must also exit this scaling solution in
order to describe the late accelerated solution. Models with
interaction between the dark sectors can account for the exit of
scaling, by dynamically modifying the scalar field potential, and
satisfy all the requirements to describe a viable dark energy scenario
\cite{dynamic_DE_DM_1,dynamic_DE_DM_2}.

 Another method used to obtain exact solutions in cosmology is the
 First Order Formalism (FOF), developed by Bazeia {\it
   et.al.}\cite{bglm}, where cosmological solutions with dark energy
 modeled by a single scalar field were obtained exactly. The central
 point of that formalism is to assume the Hubble's factor to be a
 function of the scalar field $\phi$, thus reducing the second order
 equations to first order ones. In another work Bazeia \emph{et. al.}
 \cite{bazeia2} generalized the first order formalism describing a
 universe filled with dust and dark energy. It is interesting to point
 that, in their work, a coupling between dark matter and dark energy
 arised naturally. The authors argued that this interaction is an
 effect of the applied first order formalism.

The aim of this work is to apply the FOF for a model where a
interaction between the dark sectors is present. We consider a
universe filled only with dark matter and dark energy which means that
our models only aim to model the late universe evolution.  We have
generalized Bazeias's cosmological models by explicitly adding a
fermionic field $\psi$, which plays the role of dark matter, and an
interaction term between the fermionic and the scalar parts. The
imposition of such interaction renders a more involved Friedmann
equations and equations of motion, increasing the number of variables
of the problem and reducing the constraints of the system. Since we
have this extra freedom in the model, we impose some restrictions to
the scale factor, via an {\it Ansatz} that relates it to the scalar
field itself.

Even though these restrictions may reduce the possibilities of
expansion of the universe, we show in this paper that we can still
construct a large class of exact solutions by applying the FOF for the
interacting models with the restriction of the solution for the scalar
field as a function of time, $\phi(t)$, to be invertible. This aspect
is actually essential for solving the equations exactly and is taken
for granted in much of the literature on this subject. A few works
involving dynamical analysis discard this requirement, but they are
able to find exact solutions after making some different restrictions
and approximations based on the structure of the phase space for the
models \cite{dynamic_DE}.

The solutions obtained present only accelerated, and decelerated and
accelerated periods. This is in accordance with the phase space
analysis, where in most cases we obtain non-viable cosmological
solutions and the scaling solution, that describes the right
cosmological evolution, is harder to find, obtaining only a few for
each model. 
We also note that, together with the dark sectors interaction proposed
as a hypothesis, the method introduces an interaction, as pointed in
\cite{bazeia2}. This happens because of the choice that the Hubble
parameter and scale factor depend only on the scalar field, what
renders an effective interaction between dark energy and dark
matter. We also account for this interaction and verify how this
influences the solutions.

The plan of this paper is as follows: in Section \ref{form} we
introduce our model of interacting dark energy and dark matter
introducing a fairly general interaction form. In Section \ref{FOF} we
present the first order formalism and present the {\it Ansatz}
necessary to reduce the order of the differential equations. In section
\ref{exact} we show the exact solutions obtained for different
parameters of the formalism. The last section shows our final remarks.

\section{Interacting Dark Energy and Dark Matter Model}\label{form}

First let us show the model we are going to study. We use the
Friedmann-Lemaître-Robertson-Walker metric with null curvature
\begin{eqnarray}
\label{01}
ds^2= dt^2-a(t)^2(dr^2+r^2d\theta^2+r^2 sin(\theta)^2d\phi^2),
\end{eqnarray}
where $a(t)$ is the scale factor. The action that describes this geometry 
and the material content is
\begin{eqnarray}
\label{02}
S=\int d^4x \sqrt{-g}\left(-\frac{R}{4}+\ell
\frac{1}{2}\partial_{\mu}\phi\partial^{\mu}\phi-V(\phi)+\frac{i}{2}
\left[\bar{\psi}\gamma ^{\mu }\nabla _{\mu
}\psi-\bar{\psi}\overleftarrow{\nabla} _{\mu
}\gamma ^{\mu }\psi\right] -M\bar{\psi}\psi +\beta F(\phi )\bar{\psi}\psi 
\right),
\end{eqnarray}
where $\phi$ is the scalar field and $V(\phi)$ its potential, $\psi$
is the fermionic field and $F(\phi)$ is an interaction.
The constants $M$ and $\beta$ are the mass of the fermionic field and the 
coupling constant, respectively and $\ell$ is a constant that can assume 
the values
$\pm 1$, if one wants the action to also accomodate the possibility of a
phantom field. The functions
$F(\phi)$ and $V(\phi)$ are not fixed in the beginning.

 We have to solve Friedmann's equations, the equation of motion for
 the scalar field and Dirac equation. We know that the homogeneous
 Dirac equation in a homogeneous and isotropic spacetime has a simple
 solution in terms of the scale factor and we can see that
 $\bar{\psi}\psi=\alpha/a^{3}$, where $\alpha$ is a constant related
 to the energy density of the dark matter at the present phase of the
 history of the universe \cite{micheletti}. Therefore, the system of
 equations that remains to be solved is
\begin{eqnarray}
\label{03}
\ddot{\phi}&+&3H\dot{\phi}+\frac{V^{'}}{\ell}=\frac{\beta
  F^{'}\bar{\psi}\psi}{\ell},\\
\nonumber\\
\label{05}
H^{2}&=&\frac{8\pi}{3M_{p}^{2}}\left\{\ell\frac{\dot{\phi}^2}{2}+V(\phi)+[M-\beta
  F(\phi)]\bar{\psi}\psi\right\},\\
\nonumber\\
\label{06}
\dot{H}&=&-\frac{4\pi}{M_{p}^{2}}\left\{\ell\dot{\phi}^2+[M-\beta
  F(\phi)]\bar{\psi}\psi\right\},
\end{eqnarray}
where $H=\frac{\dot{a}}{a}$ is Hubble's factor and $M_{p}$ the
Planck's mass. A prime represents derivative with respect to the
scalar field $\phi$.

We also evaluate the energy-momentum tensor. We evaluate it separately
for each component, the scalar field and the fermionic field, and for
the interaction term \footnote{Similar separation can be seen in
  \cite{Lepe}}. We then obtain the energy density, pressure and
equation of state in the Friedmann-Lemaître-Robertson-Walker
space-time. As the interaction term, $\mathcal{L}_{int}=\beta F(\phi
)\bar{\psi}\psi $, depends on both the scalar field and the fermionic
field we choose not to include the interaction as a part of any of
those fluids. Instead, we separate it as a different "component" so we
can see how it evolve and how dominant this term is, since $F(\phi)$
is not known and the method can introduce new interactions.

The energy density $\rho$, the pressure $p$
and the equation of state $\omega$ for the scalar field are given by:
\begin{eqnarray}
\rho _{DE} =\frac{\ell\dot{\phi}^2}{2}+V(\phi)\,, \qquad
p _{DE}=\frac{\ell\dot{\phi}^2}{2}-V(\phi)\,, \qquad
w_{DE} =\frac{p}{\rho}\,,
\end{eqnarray}
where for the standard field case $\ell=1$ and for the phantom field
case $\ell=-1$, leading to a change in the sign of the kinetic term
in the energy density and pressure and resulting in an equation of state
parameter such that $w \leqslant -1$. 
For dark matter we have
\begin{align}
\rho _{DM}= M\frac{\alpha}{a^{3}}\,, \qquad p_{DM}=-\beta \phi
\frac{\alpha}{a^{3}}\,, \qquad \omega _{DM}=\frac{-\beta \phi}{M}\,.
\end{align}
We can see that in the case with interaction the pressure of dark
matter does not vanish for all times but it depends on both fields and
the interaction constant. This happens because when we have
interaction the components do not conserve alone, but the sum of all
components is what is conserved. However, we can stress that depending
on the solutions, $p_{DM}$ and $\omega _{DM}$ can vanish for a certain
time. It is important to notice that this choice of separation of the
components breaks the standard distinction between the dark matter and
dark energy, since now both components can evolve as matter, dark
energy or both, depending on the solution obtained. This is
fundamental to understand the mixed role played by each component in
the solutions obtained in Section \ref{exact}.

For the interaction component, the energy density, pressure and
equation of state are given by

\begin{align}
\rho _{int}=-\beta \phi \frac{\alpha}{a^{3}}\,, \qquad p_{int}=\beta
\phi \frac{\alpha}{a^{3}}\,, \qquad \omega_{int}=-1\,.
\end{align}
The constant equation of state equal to $-1$ shows us that the
interaction component also makes the role of accelerating the
universe, as the dark energy does. Therefore, if the interaction is
too strong in comparison with the other components, this would act to
accelerate the universe.

We can also define, as usually, the density parameter $\Omega =\rho
/\rho _{crit}$, where $\rho _{crit}=3H_{0}^{2}\Gamma /2$ with $\Gamma
= M_{p}^{2}/4\pi$.

\vspace{0.5cm}

The above system of equations (\ref{03}-\ref{06}) describes the
dynamics of the variables of our interacting system. We must solve
this system in order to obtain the evolution of our components.  We
can see that the interaction term brings an explicit dependence on the
scale factor in the equations adding a new dynamical variable in the
differential equations. We can notice, for example, that the Friedmann
equation (\ref{05}) can no longer be used as a constraint equation, as
it is in the case without interaction.

Because of these differences the system is more complicated to be
analysed. If one wants to compute the solutions numerically one needs
to perform a phase space analysis, that can only be done for a power
law-like potential in the case of Yukawa couplings \cite{futuro},
requiring the imposition of assumptions in the variables to be studied
for general potentials. Because of these difficulties, we explore a
different route to study this model based on \cite{bglm,bazeia2},
where we rely on important simplifications to obtain exact
solutions. We want to check if this method and these simplifications
can be used and give good results in the context of interacting
models.

\section{First Order Formalism - FOF}
\label{FOF}

Now, we are going to introduce the formalism that we are going to
use. Following the central idea of FOF we make the simplifying
assumption that:
\begin{eqnarray}
\label{07}
H(t)=W(\phi(t))\qquad \Longrightarrow \qquad \dot{H}=W_{\phi}\ \dot{\phi}\;,
\end{eqnarray}
where the dot stands for time derivative and $W_{\phi}\equiv\partial W
/\partial \phi$. We require that $\phi$ must be invertible, so
solutions will only be possible through this method if the scalar
field and its time derivatives are smooth, monotonic functions
\cite{courant}. Other approaches can present solutions that are not
invertible (see, e.g., \cite{saa,elizalde}), but they are usually
valid only for very specific regions of the phase space of the
cosmological models. This invertibility requirement restricts the
solutions that can be obtained by the method.

With this in mind we rewrite the
Eqs.(\ref{03}-\ref{06}) substituting $H$ by
$W(\phi)$. The Eq. (\ref{06}) becomes
\begin{eqnarray}
\label{08}
-W_{\phi}\ \dot{\phi}\ \Gamma=\ell\dot{\phi}^{2}+[M-\beta
  F(\phi)]\frac{\alpha}{a^3},
\end{eqnarray}
with $\Gamma=\frac{Mp^{2}}{4\pi}$. By inspection of the Eq.(\ref{07}),
we see that it is necessary to impose some restrictions on the form of
the scale factor. In order to solve this equation in terms of
$\dot{\phi}$, we make an extra assumption and choose to rewrite the
scale factor using the {\it Ansatz}
\begin{eqnarray}
\label{09}
a(t)^{-3}=\sigma \ \dot{\phi}^{n}\ J(\phi),
\end{eqnarray}
where $\sigma$ is a real constant, $n$ is an integer and $J(\phi)$ an
arbitrary function of the scalar field. This expression has a general
form in a way that allow us to obtain a large class of exact solutions
with interacting dark energy and dark matter, by choosing convenient
integers $n$ and functions $J(\phi)$ that reduce the order of the
equations of motion. A more direct approach, without assuming this
\textit{ansatz}, can also be done as shown in Appendix 1.

Thus, substituting (\ref{09}) in (\ref{08}) we obtain
\begin{eqnarray}
\label{10}
\dot{\phi}^{n-1}&+&\frac{[\ell\dot{\phi}+W_{\phi}\Gamma]}{[M-\beta
    F(\phi)]\alpha\sigma\ J(\phi)}=0\;.
\end{eqnarray}
The potential $V_{n}(\phi)$ associated to the scalar field is
calculated using (\ref{05}), resulting in
\begin{eqnarray}
\label{11}
V_{n}(\phi)=\frac{3\Gamma W^{2}}{2}-\ell
\frac{\dot{\phi}^2}{2}-[M-\beta F(\phi)]\alpha\sigma \dot{\phi}^{n}
J(\phi).
\end{eqnarray}
We can solve Eq.(\ref{10}) as an algebraic equation for $\dot{\phi}$
for each value of $n$. Some of the roots of this equation can be
imaginary, and will be discarded. We also restrict the values for $n$
in our calculation to $n=0,\, 1,\, 2,\, ...$.  Such a procedure
reduces the order of the equations that need to be solved and
transforms the equation of motion for the scalar field into a
constraint equation. The functions $W(\phi)$, $J(\phi)$ and $F(\phi)$
need to satisfy the constraint equation:
\begin{eqnarray}
\label{13}
3W\dot{\phi}&+&\frac{3\Gamma W W_{\phi}}{\ell}-\frac{[ M - \beta
    F(\phi)]\alpha\sigma}{\ell}\frac{d}{d\phi}\left[\dot{\phi}^{n}J(\phi)\right]=0\;,
\end{eqnarray}
where we have used Eqs(\ref{10},\ref{11}) and the relation $
\ddot{\phi}=\dot{\phi}\ d\dot{\phi}/d\phi$ valid whenever the function
$\phi(t)$ is invertible. This is the point where it is crucial for
this function to be a one to one map, what is not always true.

As we can see, the power of this method is that you obtain two
important quantities. The scale factor that is exactly what it is not
known and we would like to obtain (to verify, for example, that a
scalar field with a potential yields acceleration) and not to choose
(as it is done in the direct approach); and the form of the
interaction, parametrized by $F(\phi)$\footnote{Although we put a
  general function in the interacting term that will cover a great
  range of possible interactions, we are already restricting that the
  interaction in the fermionic sector depends on $\bar{\psi}\psi$. If
  $F(\phi)=\phi$ we recover the Yukawa coupling. } , the most
important feature of the interacting models that is in principle
unknown. You only impose the form of the Hubble parameter and that the
scale factor depends on the scalar field given the form of this
dependency. The method then provides you the form of the evolution of
the system as well as the interaction between dark matter and dar
energy

So, the setup to obtain the solution is the following. Given $W(\phi)$
and $J(\phi)$ and chosen a value for $n$, we write the potential
(\ref{05}) with the chosen form for these functions and solve the
constraint equation (\ref{06}) to obtain the form of the coupling
$F(\phi)$. We plug this in (\ref{03}) and solve this algebraic
equation with the order given by the value chosen for $n$. With this
it is possible to obtain $\phi (t)$ and $a(t)$ and evaluate the energy
density and pressure of both components.

\vspace{0.5cm}

As an example, we can show the form of the equations for $n=0$. The
first order equation will be given by
\begin{eqnarray}
\label{15}
\dot{\phi}=-\frac{W_{\phi} \Gamma}{2\ell}\pm
\frac{\sqrt{\left(W_{\phi} \Gamma\right)^2-4\ell[M-\beta
      F(\phi)]\alpha\sigma J(\phi)}}{2\ell}
\end{eqnarray}
\noindent with the potential $V_{0}$ being
\begin{eqnarray}
\label{16}
V_{0}(\phi)=\frac{3\Gamma W^2}{2}-\ell\frac{\dot{\phi}^2}{2}-[M-\beta
  F(\phi)]\alpha\sigma J(\phi)\;.
\end{eqnarray}
The constraint equation relating the functions $J,W$ and the
interaction $F$ is given by
\begin{eqnarray}
\label{17}
3W\dot{\phi}+\frac{3\Gamma W_{\phi}W}{\ell}-\frac{[M-\beta
    F(\phi)]\alpha\sigma}{\ell}\frac{d}{d\phi}\left(J(\phi)\right)=0\;,
\end{eqnarray}
with the scale factor being given by $a(t)^{-3}=\sigma J(\phi)$.

As a check for this method when we choose the function $J(\phi)$ to be constant (say,
when $J(\phi)=1$), we can see that these equations represent the special case of a
Minkowski universe, known to be a fixed point in the phase space
describing the dynamics of the system.

Another important check is to see if the case of an universe with an
exponential expansion is recovered when we set $W(\phi)=const$, as it
is expected since
\begin{eqnarray}
H(\phi)=W(\phi)=H_0\quad \longrightarrow \quad \frac{\dot{a}}{a}=H_0 \quad
\longrightarrow \quad a(t)=a(0)\re^{H_0 t}\,,
\end{eqnarray}
but without imposing a form for $a(t)$. We will try to find valid
cosmological solutions using FOF for different values of $n$ and
different forms of $W(\phi)$ and $J(\phi)$.

\section{Exact Solutions for the Interacting Dark Energy and Dark Matter Models with FOF}
\label{exact}

In this section, we search for cosmologically viable solutions using the
FOF. We show some examples of solutions giving different evolutions
and couplings between the dark matter and dark energy. We intend to
study how to obtain the cosmological solutions for different couplings
and understand how well the method can describe the coupling
scenario. As in the case of the phase space analysis we expect to find
at least one cosmologically viable solution presenting a period where
the dark energy energy density becomes subdominant. In all cases
$\ell=1$.

After finding the solutions by the method, the free parameters of each
of them must be constrained by the observational data in order to
determine in which circunstances the solutions are viable. Some
solutions required the introduction of parameters in order to have the
right dimensions, that were also matched with the observations.

In the first two examples we describe an accelerated universe, given
$W=const.$, and see how the FOF for different $n$ and different
$J(\phi )$ gives us different evolution and couplings between the dark
energy and dark matter. This also serves as a test for the
\textit{ansatz} since we know that for that $W(\phi)$, the scale
factor must evolve exponentially. We are interested in verifying if
the solutions obtained are cosmologically viable by evaluating the
cosmological parameters for each example.

\vspace{0.5cm}

\subsection{$1^{st}$ Example:}

\vspace{0.3cm}

Considering the case with $n=0,\sigma=1$ and $W(\phi)=H_0$, $J(\phi)=\frac{-\phi+M}{C}$,
where C is a constant with dimension of energy ($MeV$), the
Eqs.(\ref{09}-\ref{13}) result in:
\begin{align}
\label{phi_1}
&F(\phi)=\frac{M}{\beta}+\frac{9H_{0}^2 C(\phi -M)}{\alpha\beta}\; , \qquad
V_{0}(\phi)=\frac{3}{2}\Gamma
H_{0}^2+\frac{9H_{0}^2}{2}\left(\phi -M \right)^2\;, \\ 
&\phi(t)=M -D \re^{-3H_{0}t}\; , \qquad \qquad
a(t)= \left( \frac{C}{D} \right) ^{1/3}\re^{H_{0} t}.
\end{align}

Such a universe develops a de Sitter expansion, therefore an
accelerated expansion.  We can also see this by calculating the
acceleration parameter $q=\ddot{a}a/\dot{a}^{2}$. In this case $q$ is
constant and equals to 1.  The interaction, described by $F(\phi )$
and that also appears in $V(\phi)$, accounts for the interactions that
were introduced between the dark sectors and the one introduced by the
FOF, as pointed out in \cite{bazeia2}.

Here, the Hubble constant $H_{0}$ and the $\alpha$ parameter play the
role of a coupling constant between dark matter and dark energy
through the relation $\beta_F=-\frac{9H_{0}^2C}{\alpha}$. These
quantities also redefine the mass of the dark matter as
$m_F=-\frac{9H_{0}^2 M}{\alpha}$, making it tachyonic.

In order to understand the role of such an extra interaction, the
cosmological model obtained and to confirm the accelerated expansion
seen in the acceleration parameter, we evaluate the energy density,
the pressure and the equation of state parameter for the scalar field,
\begin{eqnarray}\label{rho_1}
\Omega_{DE}(a) =1+\frac{6}{C^{2}\Gamma}\frac{1}{a^{6}} ,\qquad
p_{DE}(a) = -\frac{3}{2}\Gamma H_{0}^2,\qquad
w_{DE}(a) = \frac{-\frac{3}{2}\Gamma
  H_{0}^2}{\left[\frac{9H_{0}^2C^{2}}{a^{6}}+\frac{3}{2}\Gamma
    H_{0}^2\right]}\quad.
\end{eqnarray}
We can also obtain these quantities for the dark matter,
\begin{equation} \label{rho_1_dm}
\Omega _{DM}(a)=\frac{2\alpha M}{3H_{0}^{2} \Gamma} \frac{1}{a^{3}}\,,
\qquad p_{DM}(a)=-\frac{9H_{0}^{2}}{C^{2}}\frac{1}{a^{6}}-\frac{\alpha
  M}{a^{3}} \,, \qquad \omega _{DM}(a)=-1-\frac{9CH_{0}^{2}}{\alpha
  M}\frac{1}{a^{3}}\,,
\end{equation}
and for the interaction

\begin{align}
\Omega _{int}(a)=-\frac{6}{C^{2}\Gamma}\frac{1}{a^{6}}-\frac{2}{3}
\frac{\alpha M}{H_{0}^{2} \Gamma}
\frac{1}{a^{3}}=-\frac{p_{int}(a)}{\rho_{crit}}\,, \qquad \omega
_{int}(a)=-1\,.
\label{rho_1_int}
\end{align}
We can readily see that $\Omega _{tot}=\Omega
_{DE}+\Omega_{DM}+\Omega_{int}=1$ as given by the Friedmann equation
(\ref{05}).

Given the form of the solution, we need to determine $\alpha M$ and $C$ by
adjusting the solution to the observational
data. We constrain some of the constants by fixing the density
parameter of the dark matter today- $a_{0}=1$, the value of the scalar
factor today- to be approximately $0.26$, according to WMAP-7 data
\cite{wmap7}. This fixes $\alpha M$.

We still need to constrain $C$. Since $C^{2}$ that appears in the
energy density and in the equation of state cannot be negative so the
scale factor is not imaginary, we can see from the expression for
density parameter of dark energy that this will always be greater than
one. Adjusting it to be as close to one as possible, so we do not have
an over-density of dark energy, we were able to constrain $C$ to a
value of the order of $10^{-8}$.

\begin{figure}[!htb]
\centering
\includegraphics[scale=0.80]{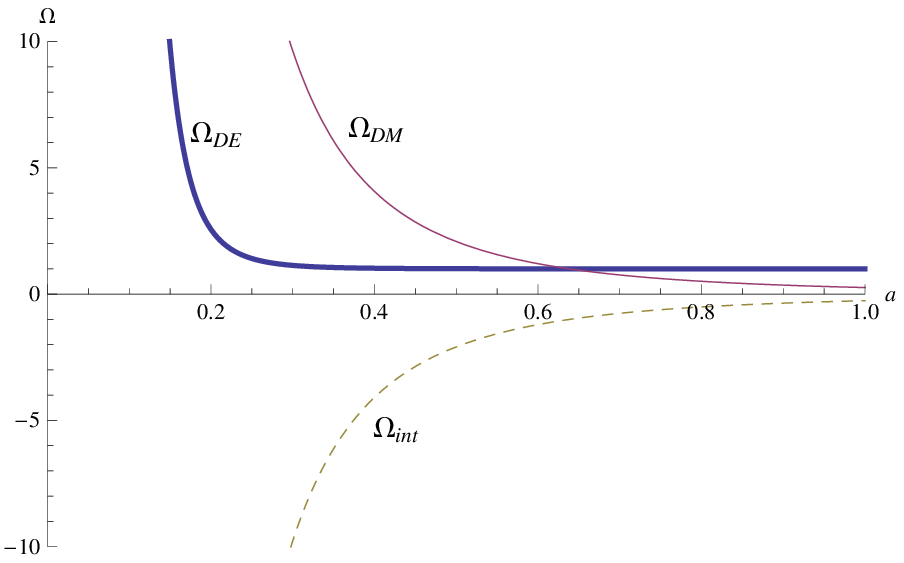}
\hspace{0.3cm}
\includegraphics[scale=0.80]{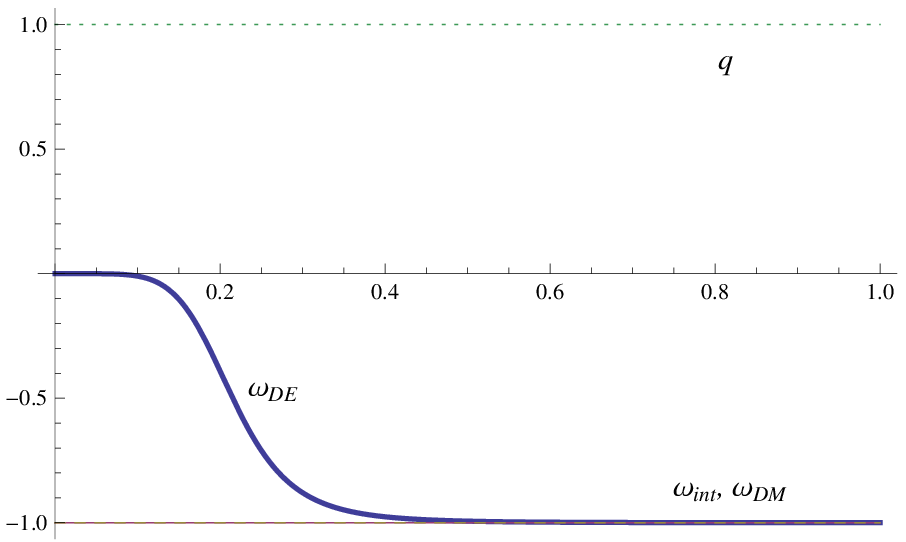}
\caption{Density parameter $\Omega$ (left panel) and equation of state
  parameter $w$ and acceleration parameter (right panel) for the model
  represented by eqs. \ref{rho_1}-\ref{rho_1_int}.}
\label{solution1}
\end{figure}

In the left panel of Figure \ref{solution1} we plot the density
parameter of all the components.  The interaction density is always
negative and large. We can see that if we put this component together
with any of the other components, how it is usual to do, this could
influence a lot the final energy density of the components.

We can see in the plot that the component that the "dark matter"
dominates in the past, while dark energy dominates at late
times. However, this does not mean that we observe deceleration in the
past, as we already know it does not happen because of the
acceleration parameter equal to $1$. As we can see in the right panel
of Figure \ref{solution1} the component we are calling dark matter
actually accelerates the universe since its equation of state
asymptotes $-1$ towards the present, instead of being equal to zero as
expected for dust. This happens for two reasons: first, because the
mass term of dark matter is cancelled by the first factor in $F(\phi)$
in (\ref{rho_1}) and only the kinetic part plays a role; second,
because of the \textit{ansatz} chosen, the scalar field will play a
role in all the terms. So, the evolution of the "dark matter"
component contains the interaction and behaves differently than the
usual dark matter.

For the dark energy component we can see that the equation of state
presents an initial period where this fluid would not accelerate the
universe, because it is larger than $-1/3$. So, this equation of state
tends to a "matter" one in the past and then, around $a=0.2$, it
enters in the regime where it accelerates the universe going to $-1$
in the present and behaving like dark energy. This feature of the
equation of state can be found in the equation of state of a Chaplygin
gas \cite{chap}, where one component can mimic the behaviour of matter
and dark energy and it is an alternative to the quintessence scenario
of dark energy \footnote{Chaplygin gas can also develop a period of
  phantom evolution as seen in \cite{chap} that we do not observe
  here.} .

However, as the dark energy component is not dominant during the
period where it does not accelerate the universe, period in which the
"dark matter" component that accelerates is dominant, the effective
behaviour is of an accelerated universe during all times.  This can be
seen also be the acceleration parameter, in the right panel of Figure
\ref{solution1}. It is important to notice that this solution
describes a flat universe (the sum of the density parameter of the
components is equal to one) that is accelerating.

Because this solution describes a universe that is only expanding in
an accelerated way, this solution is not a good candidate for a viable
cosmological solution for the late universe. The fact that we could
only obtain an accelerated solution should not be used as a reason to
discard it, since we imposed it by setting the Hubble parameter to be
constant as an hypothesis. This solution could be a good one to
describe a de Sitter inflationary period \cite{Guth} with two
different components responsible for the acceleration in different
times of the evolution. This solution is an attractor solution in the
phase space analysis of dark energy \cite{dynamic_DE} and appears
naturally in our formalism, as would have been expected.

We now show another solution, for a different $n$ but with the same
constant Hubble parameter.

\vspace{0.5cm}

\subsection{$2^{nd}$ Example:}

\vspace{0.3cm}

\begin{figure}[!htb]
\centering
\includegraphics[scale=0.80]{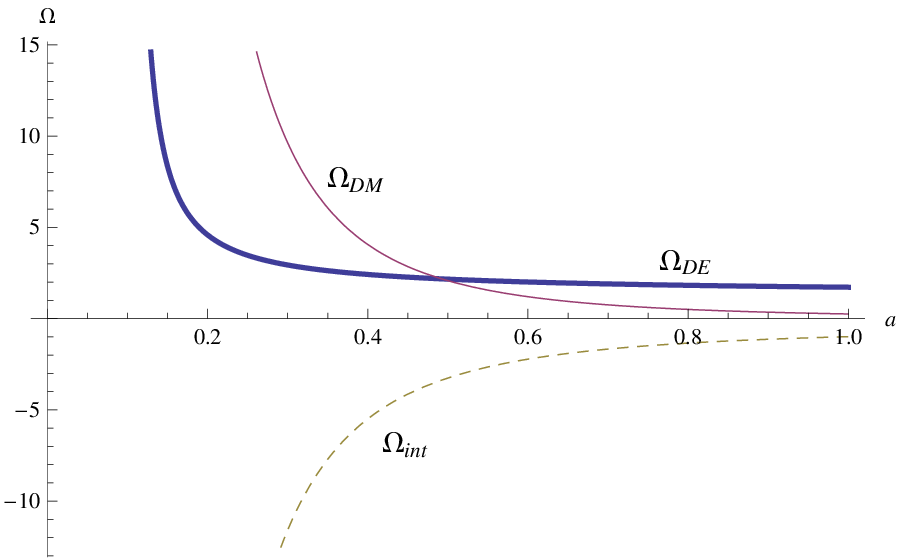}
\hspace{0.3cm}
\includegraphics[scale=0.80]{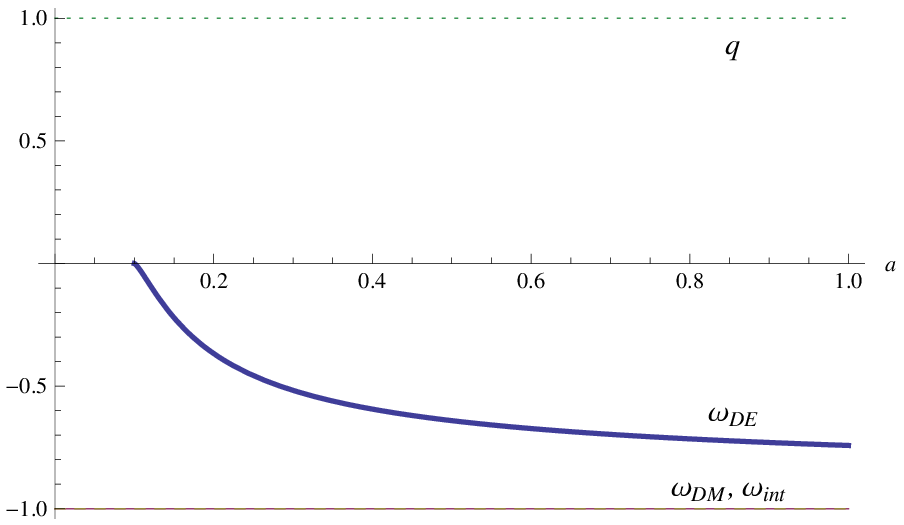}
\caption{Density parameter $\Omega$ (left panel) and equation of
  state parameter $w$ and acceleration parameter (right panel) 
for the model represented by
  Eqs.(\ref{rho_2}-\ref{rho_2_int}) .}
\label{solution2}
\end{figure}

Considering the case with $n=1,\sigma=1$ and functions $W(\phi)=H_{0}$ and
$J(\phi)=-\frac{P}{\alpha\phi^2(M-\beta F(\phi))}$, where $P$ is a
parameter with dimension 4 in energy ($MeV^{4}$), Equations
(\ref{09}-\ref{13}) result in
\begin{align}\label{phi_2}
F(\phi)=\frac{M}{\beta}+\frac{C_{1}\ \re^{H_{0}\phi ^3/P}}{\phi^4}\;, \qquad
&V_{1}(\phi)=\frac{3}{2}\Gamma H_{0}^2+\frac{P^{2}}{2\phi^4}\;,\\ \nn
\phi(t)=(3Pt)^{1/3}\;,  \qquad
&a(t)=\left( \frac{\beta\alpha C_{1}}{P^{2}} \right)^{1/3}\ \re^{H_0 t}\,.
\end{align}
As in the previous example, this universe also develops a de Sitter
accelerated expansion with acceleration parameter equal to $1$. This
can be shown by evaluating the acceleration parameter that is constant
and equal to one, as can be seen in the right panel of Figure
\ref{solution2}, showing that the solution is always accelerated.

The new coupling constant is $\beta_F=C_1\beta$ and the mass of the
dark matter is cancelled by the first term of the interaction
$F(\phi)$. In this case dark matter is represented by massless
fermions, i.e., they contribute only with a kinetic term.

Another solution for $\phi(t)$ is possible but this choice results in
a contracting universe. As we are not interested in this type of
solution, (although we find it interesting that the formalism also
presents us this sort of situation) we discard it.  Here, the
influence of the dark matter is explicitly present in the scale
factor.

For this model the energy density, the pressure and the equation of
state parameter are given by
\begin{align}
\label{rho_2}
\Omega_{DE}(a)=1+\frac{2}{3\Gamma} \left( \frac{P}{3H_{0}}
\right)^{\frac{2}{3}} \frac{1}{\left[ \ln (\gamma a ) \right]
  ^{\frac{4}{3} }}\, , \qquad p_{DE}(a) =-\frac{3}{2}\Gamma H_{0}^2\,
, \qquad w_{DE}(a) =\frac{-1}{1+\frac{2}{3\Gamma} \left(
  \frac{P}{3H_{0}} \right)^{\frac{2}{3}} \frac{1}{ \left[ \ln (\gamma
      a ) \right] ^{\frac{4}{3}}}} \,,
\end{align}
where $\gamma^{3}=P^{2}/C_{1}\alpha \beta$.

We can also obtain these quantities for the dark matter and the
interaction,
\begin{align}
\label{rho_2_dm}
\Omega _{DM}(a)=&\frac{2M\alpha}{3H_{0}^{2}\Gamma} \frac{1}{a^{3}}\,,
\qquad p_{DM}(a)=-\left( \frac{PH_{0}^{2}}{3} \right)^{\frac{2}{3}}
\frac{1}{\left[ \ln (\gamma a )
    \right]^{\frac{4}{3}}}-\frac{M\alpha}{a^ {3}} \,, \nonumber
\\ &\omega _{DM}(a)=-1- \left( \frac{P}{3H_{0}} \right)^{\frac{2}{3}}
\frac{H_{0}^{2}}{M\alpha} \frac{a^{3}}{\left[ \ln (\gamma a ) \right]^{\frac{4}{3}}}\,,
\end{align}
and for the interaction
\begin{align}
\Omega _{int}(a)=-\frac{2}{3\Gamma} \left( \frac{P}{3H_{0}}
\right)^{\frac{2}{3}} \frac{1}{\left[ \ln (\gamma a )
    \right]^{\frac{4}{3} }}-\frac{2M\alpha}{3H_{0}^{2}\Gamma}
\frac{1}{a^{3}}=-\frac{p_{int}(a)}{\rho_{crit}}\,, \qquad \omega
_{int}(a)=-1\,.
\label{rho_2_int}
\end{align}

We can constrain the constants in this example, $P$, $\gamma$ and
$\alpha M$, like we did previously by fixing the density parameter of
the dark matter today to be approximately $0.26$, according to WMAP-7
data \cite{wmap7}, to fix $\alpha M$. Setting $a=1$ today, we can
constrain the coefficient $\gamma$ to be of order of $10$. As the
density parameter of dark energy is always greater than one, the
remaining constant, namely $P$, cannot be fixed using the density
parameter of today, since $P^{2}$ cannot be negative so the scale
factor is not negative. To constrain this constant we use the equation
of state of dark energy that must be negative for all times,
asymptotically approaching -1, as shown in the right panel of Figure
\ref{solution2}. However, in order to $\omega_{DE} \propto -1$ at
$a_{0}=1$, $P^{2}$ must be small. Hence, in order to impose this
condition we fine-tune $P$ to be of order of $10^{-7}$ to give us a
realistic cosmological evolution.

In the left panel of Figure \ref{solution2} we plot the density
parameter of the component that we called dark matter, which dominates
in the past, and of dark energy that dominates in the future. The
density parameter of dark energy is always greater than one but it is
compensated, as it is the dark matter one, by the interaction that it
is always negative, giving a flat accelerated solution, with $\Omega
_{tot}=\Omega _{DE}+\Omega_{DM}+\Omega_{int}=1$. This interaction is
strong and if included in any of the other components, it would
greatly affect the evolution.

As in the first example, this does not mean that we have an early
period of deceleration.  As we can see in the right panel of Figure
\ref{solution2}, the equation of state of the "dark matter" is
negative, tending to $-1$ through all the evolution. This component
accelerates the universe, different than what is expected from
dust. Again, because in this example we have a cancellation of the
dark matter mass and because the evolution is mixed with the scalar
field one, the effective behaviour is similar to the presence of a
cosmological constant.

Also as the first example, the dark energy component decelerates in
the past, starting to accelerates around $a=0.15$ as $\omega$ reaches
$-1/3$ and continuing accelerating until the present with the equation
of state tending to $-1$. This is also similar to the Chaplygin gas,
since it evolves like dust and dark energy for different times.

The effective behavior is that we only observe acceleration, as seen
in the acceleration parameter, since the "dark matter" component that
accelerates the universe dominates during the period where the dark
energy decelerates. Then dark energy dominates and continues to
accelerate the universe.

 This solution, again, is always accelerated, as expected since the
 Hubble parameter is constant.  It could describe only the period of
 an inflationary universe with two accelerating components ath
 different times, for example, and it is not a good description for
 the dark energy in the late universe.

We present now other models in the attempt to check realistic
cosmological solutions for non-constant Hubble parameters.

\vspace{0.5cm}

\subsection{$3^{rd}$ Example:}

\vspace{0.3cm}


Considering the case with $n=1,\sigma=1$ and functions $W(\phi)=\phi+H_{0}$ and
$J(\phi)=-1/C^{2}$, where $C$ is a parameter with dimension of energy
($MeV$), Equations (\ref{09}-\ref{13}) result in
\begin{align}
\label{phi_3}
&F(\phi)=\frac{3C^{2}\phi
  \left(\frac{\phi}{2}+H_{0}\right)}{\beta\alpha} +C_{1}\;,\\ \nn
V_{1}(\phi) & =\frac{1}{8C^{4}}\left\{ 4\alpha^{2}\left(-\beta C_{1}+
M\right)^{2}+C^{2}\left[4C^{2}\left(3H_{0}^{2}\Gamma-\Gamma^{2}\right)+24H_{0}\phi
  (\alpha\beta C_{1}+C^{2}\Gamma-\alpha M)\right.\right.\nonumber \\ &
  \left.\left.+12\phi^{2}
  \left(C^{2}\left(3H_{0}^{2}+\Gamma\right)+\alpha \left(\beta
  C_{1}-M\right)\right)+36H_{0}C^{2}\phi^{3}+9C^{2}\phi^{4}\right]
\right\} \quad , \\ \nn
\phi(t)&=-H_{0}+\frac{\tilde{B}}{C}\tanh\left(\frac{t\tilde{B}}{2C}\right)\quad,
\qquad \tilde{B}=\sqrt{9H_{0}^{2}C^{2}+6(-C_{1}\beta\alpha
  -C^{2}\Gamma + \alpha M)}\,,\\ \nn a(t)&= \left(
\frac{-6C^{4}}{\tilde{B}} \cosh ^{2}\left(\frac{t\tilde{B}}{2C}
\right)\right)^{1/3}\,.
\end{align}

In this case the mass of the fermions were not cancelled and the
interaction has two effective coupling constants. The acceleration
parameter is given by

\begin{equation}
q(a)=-2+\frac{3}{\delta a^{3}-1}\,,
\label{q3}
\end{equation}
where $ \delta=-\tilde{B}/6C^{4}$. This acceleration parameter can
present different behaviours depending on the value of $\delta$. This
factor will be determined by the observational constraints on the
parameters.

For this universe, the energy density for dark energy is:
\begin{align}
\Omega _{DE}(a)& =\frac{2}{\Gamma}\left[ C^{2}(2\Gamma
  -3H_{0}^{2})+2\alpha (\beta C_{1}-M) \right]-24\frac{H_{0}\alpha
  \beta C_{1}}{\tilde{B}C} \sqrt{\frac{\delta a^{3}}{\delta
    a^{3}-1}}+\frac{\tilde{B}^{2}C^{2}}{3\Gamma} \left[
  \frac{4}{\delta a^{3}-1} +C^{2} \frac{\delta a^{3}-1}{\delta
    a^{3}}\right] \nonumber \\ &+\frac{1}{3\tilde{B}^{2}\Gamma}
\left\{ \left[ 9H_{0}^{4}+12(C^{2}-1)H_{0}^{2}\Gamma -4\Gamma
  ^{2}C^{2} \right]+4\alpha ^{2} \left( M-\beta C_{1} \right)^{2}+12
\alpha H_{0}^{2} \left( M-\beta C_{1} \right) \right\} \frac{\delta
  a^{3}}{\delta a^{3}-1}\,.
\label{rho_3} 
\end{align}
where we do not show the other parameters because they are too
lengthy.

We can also obtain these quantities for the dark matter,
\begin{align} \label{rho_3_dm}
\Omega _{DM}(a)=&-\frac{2\alpha M}{C^{2} \Gamma}\frac{1}{\delta
  a^{3}-1}\,, \qquad p
_{DM}(a)=\frac{\tilde{B}^{2}}{36C^{4}}\frac{1}{\delta a^{3}}\left[
  6\alpha \beta C_{1}-9c^{2}H_{0}^{2}+\tilde{B}^{2} C^{4}
  \frac{(\delta a^{3}-1)}{\delta a^{3}} \right]\,, \nonumber
\\ &\omega _{DM}(a)=-\frac{1}{6\alpha M} \left[ 6\alpha \beta C_{1}
  -9C^{2}H_{0}^{2}+\tilde{B}^{2}C^{4} \frac{(\delta a^{3}-1)}{\delta
    a^{3}} \right]\,,
\end{align}
and for the interaction,
\begin{align}
\Omega _{int}(a)=-\frac{2}{3C^{2} \Gamma}\frac{(\delta a^{3})^{\frac{1}{2}}}{\delta a^{3} -1} \left[  6\alpha \beta C_{1}-9C^{2}H_{0}^{2}+B^{2}C^{4} \frac{(\delta a^{3}-1)}{\delta a^{3}} \right]=-\frac{p_{int}(a)}{\rho_{crit}}\,, \qquad \omega _{int}(a)=-1\,.
\label{rho_3_int}
\end{align}

\begin{figure}[!htb]
\centering
\includegraphics[scale=0.80]{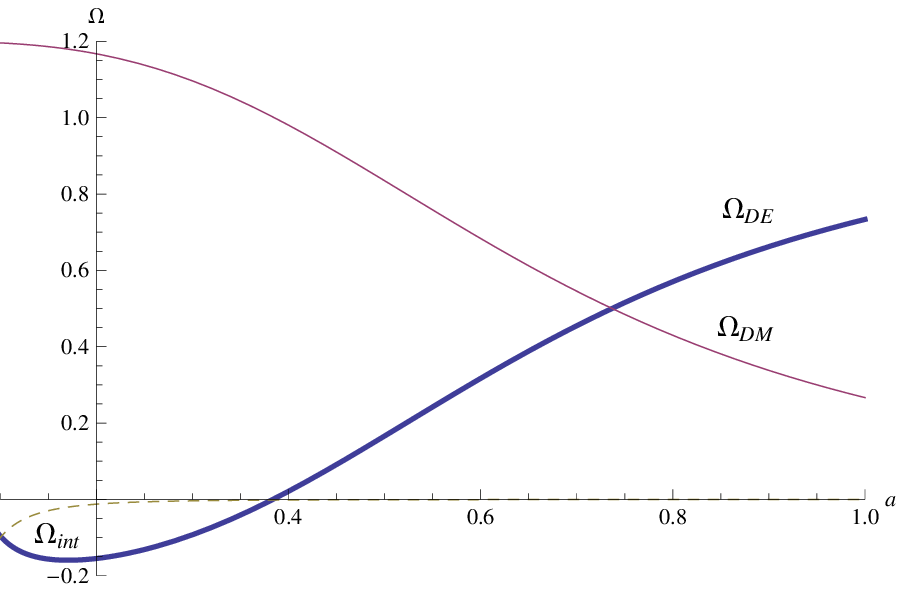}
\hspace{0.3cm}
\includegraphics[scale=0.80]{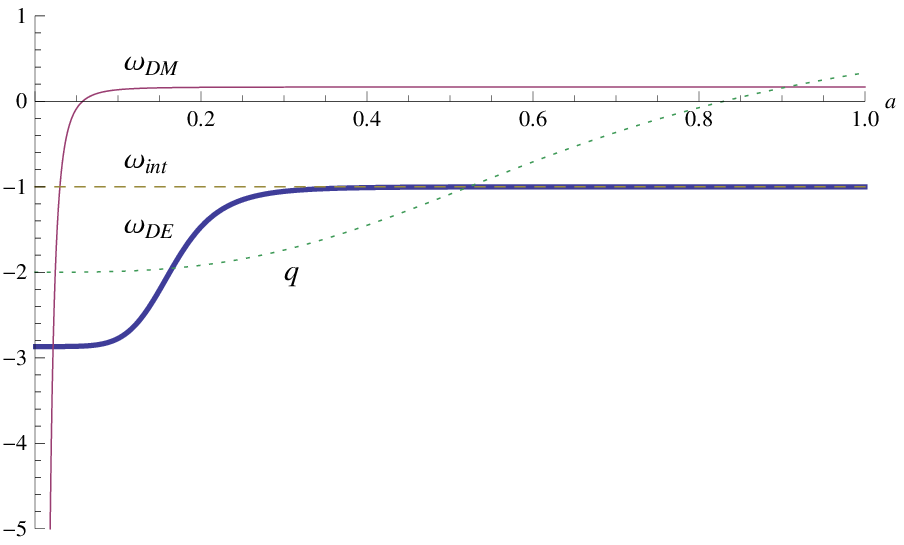}
\caption{Density parameter $\Omega$ (left panel) and equation of state
  parameter $\omega$ and acceleration parameter (right panel) for the
  model represented by Eqs.(\ref{phi_3}-\ref{rho_3_int}). The equation
  of state of dark matter is normalized in order to make the graph
  easier to see in a way still holds its main features.}
\label{solution4}
\end{figure}

We have to fix the parameters $\beta C_{1}$, $\alpha M$ and $C$ using
the observational data. From (\ref{q3}), depending on the value of
$\delta$ we can have periods of only acceleration and only
deceleration or periods that combine both. We choose a $\delta$
parameter that gives us an early period of deceleration followed by a
period of acceleration, as it can be seen in the right panel of Figure
\ref{solution4}. This is chosen because this is the expected behaviour
for the late evolution of our universe (with the restriction that here
it is only composed only by dark energy and dark matter).

To help fixing the remaining parameters, we have adjusted the density
parameter of the dark matter today to be approximately $0.26$ and the
density parameter of dark energy to be approximately $0.74$, according
to WMAP-7 data \cite{wmap7} adjusted for only two fluids.  This way we
were able to fine-tune the combinations of the constants $\beta
C_{1}$, $\alpha M$ and $C$ in (\ref{rho_3}) and (\ref{rho_3_dm}) so
the cosmological parameters agree with the observations. We plotted
the density parameter and equation of state.

In the left panel of Figure \ref{solution4} we can see that dark
matter dominates in the past, reaching its today value as $a=1$,
followed by a late period of dark energy domination, that also has the
right density value today. Like seen in the acceleration parameter in
the right panel of Figure \ref{solution4}, where we have deceleration
followed by acceleration, we can see from the equation of state that
the dark matter component decelerates the universe during all periods
while the dark energy one accelerates the universe during all
times. The only exception is during the very early universe where the
equation of state of dark matter, that is plotted normalized for
comparison reasons, is negative. However, this phase is very short and
our simplified model with only two components is not a good
approximation for this early period. The interaction density is again
always negative. However, different from the other examples it is
small compared to the others. In this case the interaction would not
spoil the behaviour of other components.

So, this model is a viable cosmological solution, since we have an
early period of dark matter domination followed by a late period of
dark energy acceleration.  It describes a flat universe since $\Omega
_{tot}=\Omega _{DE}+\Omega_{DM}+\Omega_{int}=1$, similar to the
standard cold dark matter plus a quintessence field scenario. This was
obtained by fine-tuning the free parameters of the solution. However,
the early evolution is different from this standard scenario which may
indicate that with this choice of parameters the interaction is
important only in the early universe. This shows the power of the
method to provide good candidate models to our universe.

As we can see in this example, the role of the parameters is decisive
in the behaviour of the solution. The fact that the interaction gives
large contributions in the first and second examples could be what
causes the solution to not exhibit any other behaviour than
acceleration, since the formalism gives a higher weight in the scalar
field component, by making all the assumptions on $H$ and $a(t)$ as
depending on $\phi$. This contribution is imposed by the FOF,as
pointed in \cite{bazeia2}, making the interaction large when adjusted
to the observational data and spoiling the cosmological solution. In
this third example, we obtained a good cosmological solution and the
contribution from the interaction was small.

\vspace{0.5cm}

\subsection{$4^{th}$ Example:}

\vspace{0.3cm}

We will further investigate good cosmological solutions given by the
method. We present now an example that exhibits a viable cosmological
evolution but with different features than the previous one.

\begin{figure}[!htb]
\centering
\includegraphics[scale=0.80]{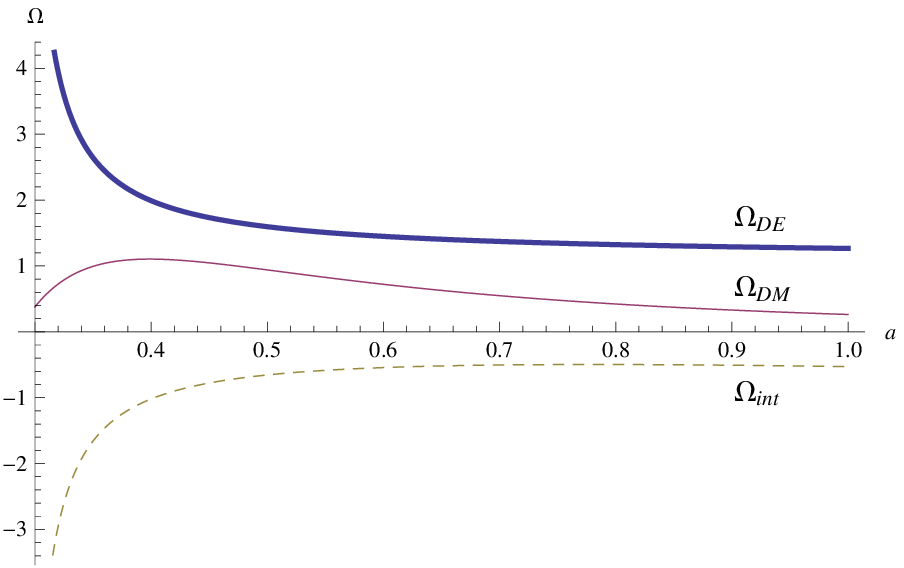} \hspace{0.3cm} %
\includegraphics[scale=0.80]{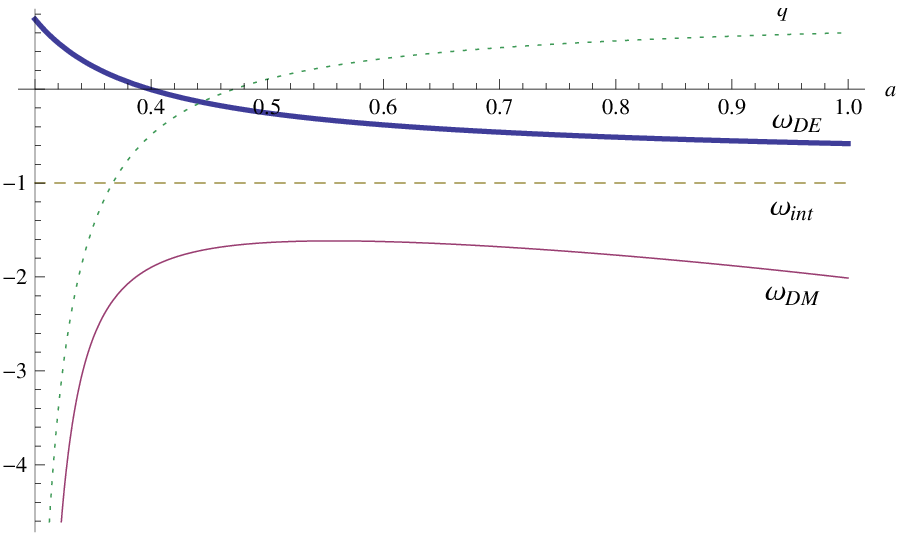} 
\caption{Density parameter $\Omega$ (left panel) and equation of
state parameter $w$ and acceleration parameter (right panel) for the 
model represented by eqs. \protect
\ref{phi_4}-\protect\ref{rho_4_int}.}
\label{solution6}
\end{figure}

Considering the case with $n=3,\sigma =1$ and functions $W(\phi )=\frac{P}{\phi \Gamma }$
and $ J(\phi )=-\frac{\phi ^{2}}{4\alpha P(M-\beta F(\phi ))}$, where
$P$ is a parameter with dimension four in energy ($MeV^{4}$), the
Eqs.(\ref{09}-\ref{13}) result in

\begin{align}
F(\phi ) =\frac{M}{\beta }+C_{1}\frac{\mathrm{e}^{\frac{3\phi ^{2}}{%
4\Gamma }}}{\phi ^{4}}  \label{phi_4}\,,  \qquad
&V_{3}(\phi ) =\frac{3P^{2}}{2\Gamma \phi ^{2}}\,, \\
\phi (t) =\left( 6Pt\right) ^{1/3}\,, \qquad
&a\left( t\right) =\left( \frac{\alpha \beta C_{1}}{2P^{2}}\right) ^{1/3}e^{%
\frac{\left( 6Pt\right) ^{2/3}}{4\Gamma }}\,.
\end{align}
This is also a case presenting massless fermionic dark matter, and an
interaction displaying a product of exponential and inverse power-law
functions, with coupling constant $\beta_{F}=C_{1}\beta$.  We can
obtain from this solution the acceleration parameter
\begin{eqnarray}
q(a)&=&1-\frac{\Gamma}{3}\left( \frac{6}{(Pt)^{2}}\right)^{1/3}\\
&=&1-\frac{1}{2\ln (\xi a)}\,,
\end{eqnarray}
where $\xi =(2P^{2}/\alpha \beta C_{1})^{1/3}$. We can see from this expression that for
\begin{equation}
t>\frac{\left( 2\Gamma \right) ^{3/2}}{6P}\text{ ,}
\end{equation}
the expansion is accelerated, given that $q>0$. So, this solution
presents a transition between decelerated and accelerated expansion as
we can see in Figure \ref{solution6}, what is expected by the
observations.

For this model the energy density, the pressure and the equation of
state parameter for dark energy are given by:

\begin{align}
\label{rho_4}
\Omega _{DE} (a)=1+\frac{1}{3\ln(\xi a)} \, \qquad
p_{DE}(a)=\frac{P^{2}}{8\Gamma ^{2}}\frac{1}{\ln(\xi a)}\left( 1-3\ln
(\xi a )\right) \,, \qquad \omega _{DE}(a)=\frac{\left( 1- 3\ln (\xi a
  )\right)}{ \left( 1+3\ln (\xi a )\right)}\,,
\end{align}
For the dark matter and the interaction,
\begin{align} \label{rho_4_dm}
\Omega _{DM}(a)=\frac{8\Gamma ^{2}M\alpha}{3P^{2}} \frac{\ln(\xi a)}{a^{3}}\,, \qquad
p _{DM}(a)=-\frac{P^{2}}{8\Gamma ^{2}} \frac{1}{(\ln(\xi a))^{2}}-\frac{\alpha M}{\xi a^{3}}\,, \qquad
\omega _{DM}(a)=-1-\frac{P^{2}}{8\Gamma ^{2} \alpha M} \frac{a^{3}}{\ln(\xi a)}\,,
\end{align}
and for the interaction

\begin{align}
\Omega _{int}(a)=-\frac{1}{3\ln(\xi a)} -\frac{8\Gamma
  ^{2}M\alpha}{3P^{2}} \frac{\ln(\xi
  a)}{a^{3}}=-\frac{p_{int}(a)}{\rho_{crit}}\,, \qquad \omega
_{int}(a)=-1\,.
\label{rho_4_int}
\end{align}

The parameters we have to fix in this example are $\xi$, $P$ and
$\alpha M$. First, we can fix the parameters by imposing that the
transition from deceleration and acceleration happens in the near
past, as expected for a good cosmological model, choosing $\xi \sim
2,02$. We also adjusted the density parameter of the dark matter today
to be approximately $0.26$, fine-tuning the ratio $\alpha M /P$.

With that we can evaluate the density parameters, plotted in the left
panel of Figure \ref{solution6}. The density parameter of dark energy
is always greater than one and dominates through all evolution. The
dark matter parameter is also plotted and is subdominant and the
interaction is always negative and is small compared to the other
components, except for very early times.

Although dark energy dominates during all times, as it can be seen by
the equation of state of dark energy in the right panel of Figure
\ref{solution6}, the dark energy decelerates the expansion in early
times. It changes from deceleration for acceleration when reaches
$\omega=-1/3$, in the same time as the acceleration parameter changes
sign. This shows that the density and equation of state evolutions are
coherent with the evolution shown in the acceleration parameter.

As already pointed out, this dark energy component that behaves like
"matter" and dark energy is analogous to the Chaplygin gas and in this
example it is responsible for the entire evolution of the universe
mimicking its components.

The so called dark matter component has a large negative value. This
happens because of the interaction and because in this example, as in
the first and in the second, the dark matter mass is cancelled.

This solution represents a flat universe, as
$\Omega_{DE}+\Omega_{DM}+\Omega_{int}=1$, presenting all the necessary
requirements for a viable cosmological model and describes a universe
where the same scalar field is responsible for the early inflationary
acceleration and the late one.

We notice that in this model the interaction given by the FOF is
small, after adjusting the parameter to fit the observational data,
and it represents a viable cosmologial solution. A possible reason for
this is that the form one chooses for $W(\phi )$ and $J( \phi )$ leads
to a coupling that can potentially spoil the cosmological solution,
since only small couplings are cosmologically acceptable in models of
interaction between dark matter and dark energy and since the FOF
favours the energy density of the scalar field and large parameters
enhance this component.

\section{Conclusions}

Interacting cosmological models have attracted much attention
in the last few years, both as a theoretical laboratory as well
as a describing phenomenology. However, due to the enormous difficulties
in obtaining solutions for the Friedmann equations in such an approach,
the theoretical effort has not gone too far. The best description of
the possible solutions to these problems comes from the dynamical
systems' analysis, which gives an overview of all the families of
possible solutions. One could ask for specific exact solutions for a particular
interaction, what is known to be very difficult, if not
completely impossible, as is usual in the framework of General Relativity.

The method described here has the purpose of simplifying the search
for exact cosmological solutions for an interacting dark sector. Such
a simplification takes place after a series of assumptions, that
however do not weaken the value of the solutions since they can be
suitable for some specific situations.

We have obtained a series of solutions and discussed their
properties. As in the phase space analysis method, we have obtained
solutions where the dark energy were the only dominant component and
two solutions where we describe both decelerated and accelerated
periods. They are viable cosmological solutions since the deceleration
period will allow for structure formation, and the accelerated one
will account for the observed late accelerated expansion of the
universe.

However, although they present the same behaviour these solutions
accomplish that in a different way. The third solution has a massive
dark matter component with an equation of state that decelerates the
universe, dominating in the past, and a dark energy component that
accelerates in the late epoch. The fourth example has a non-massive
dark matter component always subdominant and a dark energy component
that plays the hole of a matter-like component, decelerating the
universe in the past, and of dark energy, accelerating it in late
times. This behaviour is similar to the Chaplygin gas, although it
does not present a phantom epoch.

The first and second solutions present an accelerating-only
expansion. As a constant Hubble parameter has been imposed, they show
that the method is robust. The interaction given by the method plus
the imposed one are large in this examples. In the third and fourth
examples, the ones that give viable cosmological solutions, the
interaction is small, at least in most of the evolution.

Hence, we have showed that the FOF is capable of giving viable
cosmological solutions in the same way as the phase space analysis. It
also gives the expected attractor solutions where only expansion is
observed.

The analysis of specific models, with well-motivated interaction
functions and coupling constants, and the comparison with
observational data will be the subject of future work.

\begin{acknowledgments}
The authors would like to thank Alberto Saa for fruitful discussions,
and FAPESP (Funda\c c\~ao de Amparo \`a Pesquisa do Estado de S\~ao
Paulo), CNPq (Conselho Nacional de Desenvolvimento Cient\'\i fico e
Tecnol\'ogico), FAPEMIG (Funda\c c\~ao de Amparo \`a Pesquisa do Estado de 
Minas Gerais) and CAPES (Coordena\c c\~ao de Aperfei\c coamento de
Pessoal de N\'\i vel Superior) for financial support.
\end{acknowledgments}


\appendix

\section*{Appendix: Direct Approach for Solving the System of Equations}
It is worthy to stress the difference between the FOF and
the more direct approach to the solution of the cosmological
equations. We shall do this by presenting the following example.

We take $F(\phi)=\phi$ and choose the de
Sitter solution ($\dot a/a=H_{0}$). In this case the equation of motion for
the scalar field can be written as
\be
\ell(\ddot\phi+3H_{0}\dot\phi)+V'=\frac{\beta\alpha}{a^3}\;.
\ee
Using Friedmann's equations, we can eliminate the potential $V$ and
solve for the scalar field $\phi$, obtaining
\be
\phi(t)=K_1+K_{2} \re^{-3H_0t}+K_{3} \re^{-\frac{3}{2}H_0t},
\ee
where $K_1$, $K_2$ and $K_3$ are constants. Note that, since there are
no restrictions on the sign of these constants, the requirement of
invertibility for $\phi(t)$ is not fulfilled for all the solutions.

For a power-law scale factor $a=K t^{p}$, with $K$ and $p$ positive
constants, we have for $\phi(t)$
\be
\phi(t)=Y_1+Y_2\left[\Frac{(\ln t)^2}{2}+Y_3 \ln t\right],
\ee
where $Y_1$, $Y_2$ and $Y_3$ are constants. This solution is clearly
non-invertible. Thus the FOF and the direct method can be understood
as complementary formalisms.


\end{document}